\documentclass{pasj01}
\usepackage{mathrsfs} 
\usepackage{graphicx} 
\usepackage{mediabb}
\usepackage{color} 
 
\title{Giant Elephant Trunks from Giant Molecular Clouds} 
\author{Yoshiaki \textsc{Sofue}\altaffilmark{} }

\altaffiltext{}{Institute of Astronomy, The University of Tokyo, Mitaka, Tokyo 181-0015\\
\\
(Appeared in {\bf PASJ, 2019, 71,121}, https://doi.org/10.1093/pasj/psz106) }

\KeyWords{ISM: molecules --- ISM: nebulae --- ISM: cloud --- stars: formation --- Galaxy: spiral arm }

\begin{document} 
\date{ } 
\maketitle  
\def\vlsr{v_{\rm LSR}} \def\Msun{M_\odot} \def\Lsun{L_\odot} \def\deg{^\circ} 
\def\r{\bibitem[]{}} \def\/{\over} \def\kms{km s$^{-1}$} 
\def\Vsun{V_0} \def\Vrot{V_{\rm rot}} 
\def\Tb{T_{\rm B}} \def\Tmb{T_{\rm mb}}
\def\sin{{\rm sin}\ } \def\cos{{\rm cos}\ } 
\def\Hcc{ H cm$^{-3}$ } \def\co{$^{12}$CO$(J=1-0)$ }
\def\be{\begin{equation}} \def\ee{\end{equation}}
\def\Kkms{K \kms} \def\Ico{I_{\rm CO}} \def\Xco{X_{\rm CO}} \def\mH{m_{\rm H}}
\def\x{\times}\def\Hcc{H cm$^{-3}$} \def\fmol{f_{\rm mol}} 
\def\({\left(} \def\){\right)} \def\[{\left[} \def\]{\right]}  
\def\Kkms{ K \kms } \def\Hsqcm{ H cm$^{-2}$ } \def\hcc{{\rm H \ cm^{-3}}}
\def\mum{$\mu$m }    
\def\Mvir{M_{\rm vir}}
\def\htwocc{\htwo cm$^{-3}$ } \def\htwocol{\htwo cm$^{-2}$ } \def\htwo{H$_2$} 
\def\Htwo{H$_2$ } \def\NHtwo{N_{\rm H_2}} \def\nHtwo{n_{\rm H_2}} 
\def\Xco{X_{\rm CO}} 
 
\begin{abstract} 
We report the discovery of large elephant trunk (ET)-like objects, named giant elephant trunk (GET),  of molecular gas in star forming complexes in the Scutum and Norma arms using the \co-line survey data with the Nobeyama 45-m telescope. In comparison with the CO maps of ETs in M16 as derived from the same data, we discuss physical properties of the GETs. Their lengths are $\sim 20$ to 50 pc, an order of magnitude larger than ETs. GETs show a cometary structure coherently aligned parallel to the galactic plane, and emerge from bow-shaped concave surface of giant molecular clouds (GMC) facing the HII regions, and point down-stream of the gas flow in the spiral arms. The molecular masses of the head clumps are $\sim 10^3 -10^4 \Msun$, about 3 to 4 times the virial masses, indicating that the clumps are gravitationally stable. Jeans masses calculated for the derived density and assumed kinetic temperature are commonly sub-solar.  We suggest that the GET heads are possible birth sites of stellar clusters, similarly to ETs' globules, but at much greater scale.  We discuss the origin of the GETs by Rayleigh-Taylor instability due to deceleration of GMCs by low density gas stagnated in the galactic shock waves as well as by pressure of the HII regions. 
\end{abstract}  

\section{Introduction}

Elephant trunks (ET) provide us with unique tool to probe the interstellar physics of shock-accelerated molecular clouds at the interface with expanding HII regions as close-up views of triggered star forming (SF) fronts (Frieman 1954; Spitzer 1954; Osterbrock 1957). 
There have been a number of optical and infrared observations of ETs associated with galactic HII regions (Pottasch 1956; Hester et al. 1996; Sugitani et al. 2001, 2007; {Carlqvist, Gahm, \& Kristen} {2002}; Chaughan et al. 2012; Getman et al. 2012; Schneider et al. 2016; M{\"a}kel{\"a} et al. {2017}; Pattle et al. 2018; Panwar et al. 2018; and the literature therein), where heavy extinction has indicated that the ETs are dense molecular clouds. 

However, only a limited number of molecular line observations have been obtained, which directly measure the physical properties of ETs such as the gaseous mass and kinematics
(
 {Sherwood \& Dachs} {1976};%CO N6231 stars 
 {Schneps et al.} {1980}; %rose co globule
 {Gonzalez-Alfonso \& Cernicharo}{1994}; %CO 
 {Massi et al.} {1997}; % CO N6357=G353
 {Pound} {1998}; %M16 eagle CO BIMA
{White et al.} {1999};%M16 Eagle CO3-2
{Gahm et al.} {2006, 2013}; %rose CO rotating 
% {Gahm et al.} {2013};%CO
 {Haikala et al.} {2017}; %CO apex 18-20" reso 
 {M{\"a}kel{\"a} et al.} 2017; %rose_ir_co
 Xu et al. 2019
% Nishimura et al. 2017 %nanten2 M16 GMC
 ).

In this paper we report the result of mapping of galactic ETs and a new type of larger-sized ET-like objects, named giant elephant trunks (GET), in the \co line emission using the FUGIN\footnote{FUGIN = FOREST Unbiased Galactic Plane Imaging survey with the Nobeyama 45-m telescope; FOREST = FOur-beam REceiver System on the 45-m Telescope} CO-line survey (Umemoto et al. 2017). 

An advantage to use the \co line is the universal linear relation between the line intensity and column density of \Htwo gas through the conversion factor $\Xco$ (Bolatto et al. 2013). The relation is well established to yield an almost constant value of $\Xco\sim 2\times 10^{20}$ \htwo (K \kms)$^{-1}$ within a factor of $\sim 1.3$ for the solar abundance, and applies not only to virialized clouds (Solomon et al. 1987), but also to non-virialized low mass clouds (Sofue and Kataoka 2013).
 
On the other hand, it must be remembered that $\Xco$ is a statistical coefficient, so that it might not be accurate for individual clouds. So, we here consider that the error in the estimation of column density is about the same as the scatter in the mass-to-CO luminosity plots (Solomon et al. 1987), which is by a factor of $\sim 1.5$ for small mass clouds. It will be worth to mention that recent observations of the GMC and MCs in the W43 complex showed that $\Xco$ masses for individual clouds coinside with those from spectral analysis of the $^{13}$CO line under the LTE (local thermal equilibrium) assumption within a factor of $\sim 1.3$ (Kohno et al. 2019).  

We first present result for the Eagle nebulae (M16) as a template of typical ETs, and then, report the discovery of a new type of ETs, which are greater in size and mass than the currently studied ETs by an order of magnitude, and name it giant elephant trunk (GET). We discuss the implication of GETs to the galactic-scale star formation in the Milky Way. 
 In this paper, ET and  GET are defined as elongated cometary structures protruding from the concave surface of molecular clouds facing contacting HII regions. The names are only for the morphlogical similarity to elephant's nose, but not for the physics that covers a variety of instabilities and collapsing processes of molecular gas at the interface with the HII regions. Thus, "GET" is a scaled-up "ET" only by morphology, while the physics may not necessarily be the same as will be discussed in the last section. 

\section{Data and Reduction}
 
The FUGIN data are presented in fits-formated cubes of maps of the main-beam temperature, $\Tmb$, in the $(l,b,\vlsr)$ (longitude, latitude, local-standard-of-rest velocity) space.  
  Details of the obsevations, reduction and calibration procedures are described in Umemoto et al. (2017): the original beam width of the Nobeyama 45-m telescope was $14''$, the antenna main-beam efficiency was 0.43, and velocity resolution was 1.3 \kms. In the data cubes, the main-beam tempearature for an effective beam size of $20''$ after regridding with a grid interval of 8.5'' and velocity interval of 0.65 \kms is presented. In this paper, the brightness temperature, $\Tb$, of the sources is assumed to be represented by the main-beam temperature, $\Tb = \Tmb$. 

The column density of hydrogen molecules is obtained using the well established relation
\be
\NHtwo=\Xco \int \Tb dv,
\ee
where $\Xco=2\times 10^{20}$ \Htwo cm$^{-2}$ is the conversion factor. 
Note that, as mentioned in the previous section, the error in column estimation using $\Xco$ will include uncertainty of about a factor of $\sim 1.5$, and propagates as it is to the estimations of the density and mass, and to the discussion of virialization. 
 
The volume density is estimated by
\be
\nHtwo=\NHtwo/D_z,
\ee
where $D_z$ is the line-of-sight full width.
For the tails, we assume that the width is equal to the depth, $D_z\sim D_y$. 
For head clumps, we assume that the depth is represented by the size diameter $D$  defined through
\be
D_z=D=\sqrt{D_x D_y}.
\ee
Here, $D_x$ and $D_y$ are the full widths of half maximum of the clump in the directions along and perpendicular to the tail, respectively, after correction for the beam size of $\theta=20''$. Namely, the full width is obtained using the apparent width on the map as 
\be
D_i=\sqrt{D_i({\rm map})^2-(d\ \theta)^2}~~ (i=x, y)
\ee
with $d$ being the distance from the Sun.

The total mass of molecular gas is estimated by
\be
M_{\rm gas}=\mu \mH \NHtwo \pi \left(D(\rm map)\/2\right)^2,
\ee
where $\mu=2.8$ is the mean molecular weight and $\mH$ is the hydrogen mass.

According to Binney and Tremain (2008), %section 4.8, p361
the Virial mass of a cloud with velocity dispersion $\langle v^2 \rangle$ and half-mass radius $r_{\rm h}$ is approximately expressed as 
\be
\Mvir \sim {1\/0.45}\ {r_{\rm h}\langle v^2 \rangle \/G}.
\ee
We here assume that the beam-corrected diameter $D$ and full width of half maximum $\delta v$ of line profile are related to $r_{\rm h}$ and $\langle v^2 \rangle$ through 
$r_{\rm h}\sim D/2$ and 
$ \langle v^2 \rangle\sim (\sqrt{2}\delta v/2)^2$ with $\sqrt{2}$ being the correction for the projection effect from the radial velocity. Then, we obtain
\be
\Mvir(\Msun) \sim 0.555\ {D \delta v^2\/G}
=231\ D({\rm pc}) \delta v({\rm km\ s^{-1}})^2. 
\label{Mvir}
\ee
This may be compared with the expression by Solomon et al. (1987),
$\Mvir=({3 f_p/8}) {D \delta v^2/G}$, where $f_p$ is a projection factor.

\section{Elephant Trunks in M16}

\begin{figure*} 
\begin{center} 
\includegraphics[width=12cm]{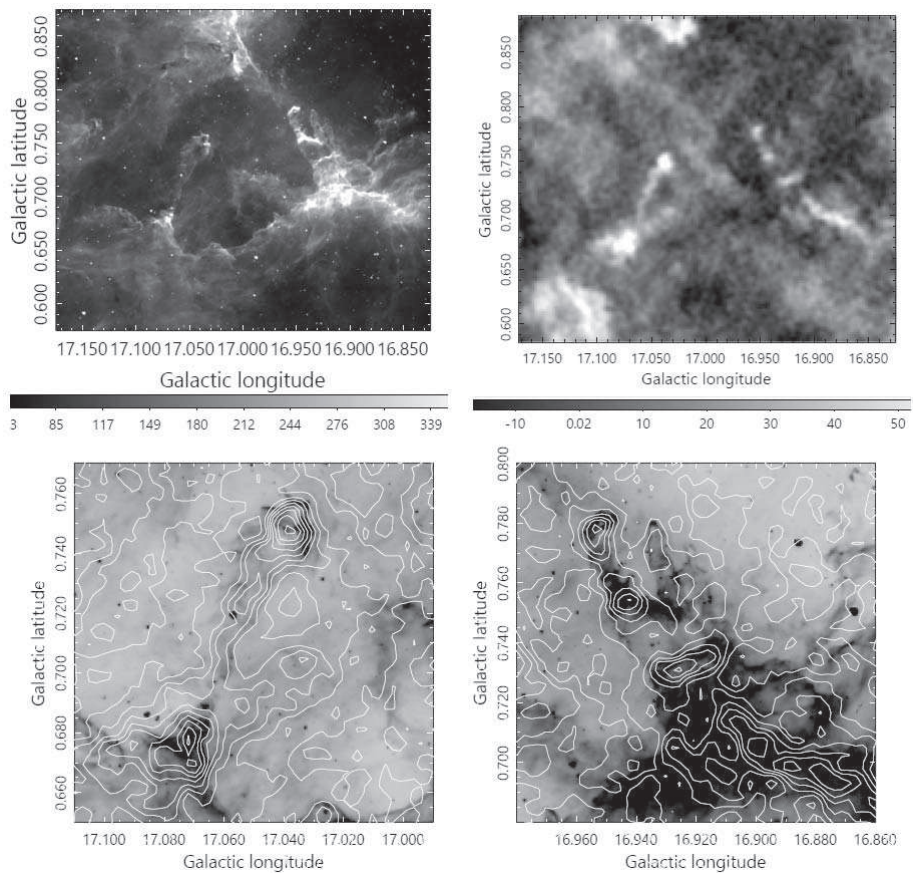} 
\end{center}
\caption{M16 ETs East and West (Eagle's pillars): (tl) 8-$\mu$m brightness map from ATLASGAL in grey scale (mJy str$^{-1}$), (tr) \co $\Ico$ (K m s$^{-1}$) without BGF, (bl) $\Ico$ by contours at 10 K \kms interval on 8 \mum for ET M16 East, and (br) West. } 
\label{M16IR} 
 	\end{figure*}

The search for GETs was performed referring to properties of ETs in M16 as a template. 
Figure \ref{M16IR} shows a BGF (background filtered) intensity map integrated from $\vlsr=20$ to 30 \kms around M16's elephant trunks comapred with the 8 \mum map from ATLASGAL (Churchwell et al. 2009). . 
ET M16 East's head position is at G017.04+0.75+25.0 ($l=17\deg.04$, $b=+0\deg.75$, $\vlsr=25.0$ \kms), and West Pillar I head is at G016.96+0.78+25.0. 
Both M16 E and W show cometary, head-tail structures.

\begin{figure*} 
\begin{center} 
\includegraphics[width=12cm]{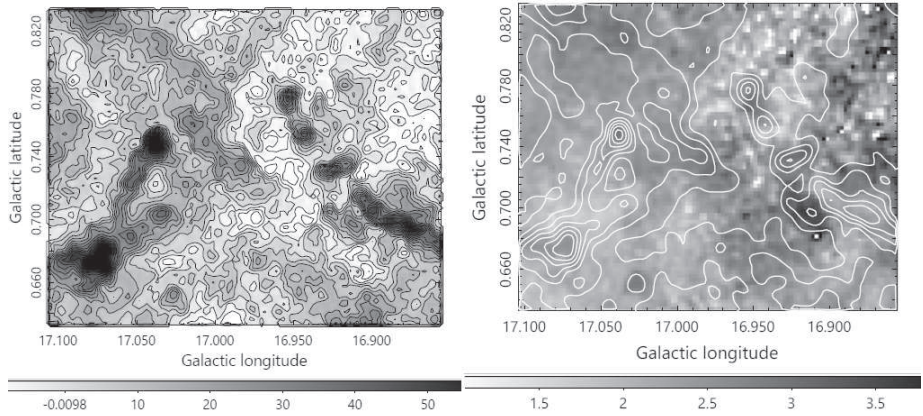} 
\end{center}
\caption{(Left) M16 ETs in \co line BGF (background-filtered) $\Ico$ map with contours at interval 5 K \kms on grey-scale map (K km s$^{-1}$), and (right) offset-velocity field (moment 1 map) in grey scale (km s$^{-1}$) overlaid by $\Ico$ contours. } 
\label{M16CO} 
 	\end{figure*}

\begin{figure} 
\begin{center} 
\includegraphics[width=8cm]{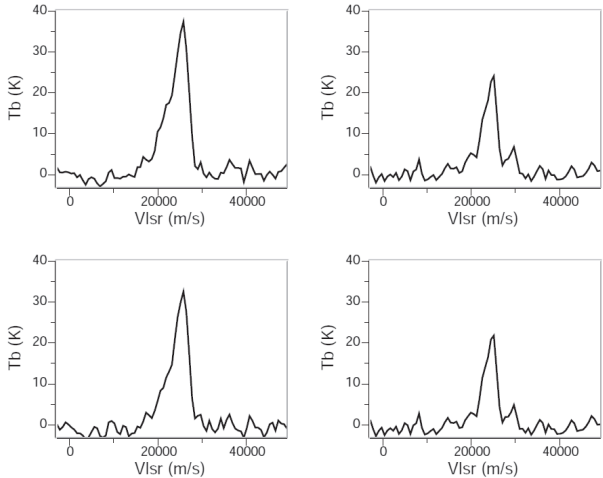} 
\end{center}
\caption{  
 (Top left) line profile (non BGF) at the head of ET M16E and (tr) M16 W, and  (bl) line profiles after BGF of ET M16E and (br) M16W. 
 } 
\label{M16COspec} 
 	\end{figure}

Figure \ref{M16CO} shows a close up view with contours every 5 K \kms. The scanning effect in the original FUGIN data has been removed using the pressing method, and the diffuse background emission has been subtracted by applying the BGF technique (Sofue and Reich 1979). See appendix \ref{BGF} for pressing and BGF techniques. Figure \ref{M16COspec} shows \co line profiles at the clump heads of M16 E and W before and after the BGF is applied.
  

The distance to the ETs has been often assumed to be equal to that of the central cluster NGC 6611, optical measurements of which showed a distance from 1.7 to 2.14 kpc (Hillenbrand et al. 1993; Guarcello et al. 2007). We here adopt a distance of $\sim 2$ kpc for the ETs. 
 
The $\Ico$ map shows remarkable similarity of the global morphology to that in infrared. However, the CO intensities in ET E and W are reversed. 
This means that the 8 \mum\ image manifests the surfaces illuminated by UV, whereas $\Ico$ map represents the molecular gas column density. 

 Another important fact is that the 8 \mum\ map contains foreground emissions, particulary in the bottom half, whereas $\Ico$ map reveals the gas properly belonging to the ET because of the restricted velocity range. This means that the lengths of the ETs of molecular gas are much longer than those recognized as pillars in optical and infrared imaging. 
 
The velocity field shows a gradient of $dv/dx\ \sin\ i\sim -0.3$ \kms pc$^{-1}$ along the major axis of M16 E, where $i$ is the inclination and $x$ is the distance from the tongue head. If the gas is accelerated from head to tail, the negative gradient indicates that the tail is nearer than the head. On the other hand, M16 W shows wavy velocity variation, possibly due to superposition of fore- and/or back-ground emissions. 

Obtained parameters are listed in tables \ref{tab1} and \ref{tab2}. Figures \ref{M16CO} and \ref{M16IR} reveal well developed long tongues for the M16 East (E) and West (W). The CO tail of West Pillar I is found to be as long as $\sim 10'$, twice that seen in optical and IR images ($\sim 5'$: Hester et al. 1996). This is thanks to the high velocity resolution of the radio line measurement, abstracting the proper structure of the ET, without being contaminated by fore- and background nebulae.

The estimated parameters for the head clump of West Pillar I (table \ref{tab2}) may be compared with the CO($J=3-2$)-line observations by White et al. (1999), who report a molecular mass of $M_{\rm mol}\sim 60\Msun$ and entire pillar (finger) $\sim 100 \Msun$. The difference by a factor of three for the clump mass will be due to our BGF procedure to subtract the diffuse component, by which the peak intensity is reduced by a factor of 0.67, as well as to the difference between employed conversion factors from intensity to column. The larger mass of the entire pillar in our estimation is obviously due to our larger (longer) area of the ET.

On the other hand, an order-of-magnitude greater values, $M_{\rm mol}\sim 300 \Msun$, $N_{\rm H_2}\sim 1.5\times 10^{23}$ \htwocol and $n_{\rm H_2}\sim 2\times 10^5$ \htwocc, are reported for the head clump of Pillar I by \co observations with BIMA (Pound 1998). 
The reason for the differences between the three observations, particularly the order-of-magnitude difference with the interferometer observation in the same line, remains unresolved here, which would be mainly due to the different ways of conversion processes from intensity to column.

Comparison with CO results with other ETs might help to judge, if our measurement is reliable. CO observations toward clumps of ETs of Rosette Nebula indicate $n_{\rm H_2}\sim 10^4$ \htwocc, $M_{\rm mol}\sim 6-11 \Msun$ (Schneps et al. 1980), and ETs in four other HII regions from $\sim 2\Msun$ to $29\Msun$ (Gahm et al. 2003). These values are comparable to the estimations for M16 ET, except for the BIMA measurement.

%Thus, we shall discuss the following results, remembering the uncertainty by a factor of three, at most ten. However, we stress that the present results on the universal $\Xco$ conversion factor for \co line has the great advantage, since the estimated values can be directly compared with that for parent MC and GMC, from which the ETs grow. 

%%%%%%%%%%%%%%%%%%%%%%%%%%%%%%%%%%%%%%%%%%%%%
	\begin{table*}  

\caption{Parameters of head clumps of ET and GET}
    \begin{center} 

\begin{tabular}{llllllllllllllllllllll} 

\hline 
\hline 
Name & $l$& $b$ &$\vlsr$& Dist. $d$ 
& Size para. $D$ &Full v. wid. $\delta v$ 
& $\Tb$ & $\Ico$\\
%& $N_{\rm H_2}$ & $n_{\rm H_2}$ 
%& $M_{\rm mol}$ & $M_{\rm vir}$ & $M_{\rm mol}\/M_{\rm vir}$ 
%& $t_{\rm J}$ & $\lambda_{\rm J}$ & $M_{\rm J}$ \\ 

& deg & deg & \kms & kpc& pc  &\kms
& K & K \kms \\
%& \htwocol & \htwocc 
%& $\Msun$ & $\Msun$ &
%& My & pc & $\Msun$ \\ 

 \hline  

M16E & 17.04& 0.75& 25& 2 &   0.36&   3.5&  40&  77\\
%& 0.15E+23& 0.14E+05& 0.46E+02& 0.71E+02& 0.64& 0.14& 0.10& 0.55\\

M16 W & 16.96& 0.78& 25& 2 &   0.27&   3.5&  25&  50\\
%& 0.10E+23& 0.12E+05& 0.19E+02& 0.52E+02&  0.36&  0.15&  0.09& 0.29\\
\hline
G31 & 30.99& -0.05& 83& 5.5 &  1.00&  3.0& 40& 120\\
%& 0.24E+23& 0.78E+04& 0.54E+03& 0.14E+03&  3.73&  0.18&  0.14& 0.73\\
G24.6 & 24.63& 0.18& 116& 7.3 &  4.15&  3.2& 10& 30\\
%& 0.60E+22& 0.47E+03& 0.19E+04& 0.68E+03&  2.73&  0.74&  0.28& 0.37\\
G24.8 & 24.80& 0.10& 111& 7.3 &  5.13&  7.0& 17& 160\\
%& 0.32E+23& 0.20E+04& 0.15E+05& 0.40E+04&  3.77&  0.36&  0.18& 0.40\\
\hline
\\
\hline 
\hline 
Name %& $l$& $b$ &$\vlsr$& $d$ 
%& $D$ &$\delta v$ 
%& $\Tb$ & $\Ico$ ($^*$)
& $N_{\rm H_2}$ & $n_{\rm H_2}$ 
& $M_{\rm mol}$ & $M_{\rm vir}$ & $M_{\rm mol}/M_{\rm vir}$ 
& $t_{\rm J}$ & $\lambda_{\rm J}$ & $M_{\rm J}$ \\ 
% deg & deg & \kms & kpc& pc &\kms
%& K & K \kms
& \htwocol & \htwocc 
& $\Msun$ & $\Msun$ &
& My & pc & $\Msun$ \\ 

 \hline 

M16E %& 17.04& 0.75& 25& 2 &  0.36&  3.50& 40.00& 77.00
& 0.15E+23& 0.14E+05& 0.46E+02& 0.71E+02&  0.64&  0.14&  0.10&  0.55\\
M16 W %& 16.96& 0.78& 25& 2 &  0.27&  3.50& 25.00& 50.00
& 0.10E+23& 0.12E+05& 0.19E+02& 0.52E+02&  0.36&  0.15&  0.09&  0.29\\
\hline
G31 %& 30.99& -0.05& 83& 5.5 &  1.00&  3.00& 40.00& 120.00
& 0.24E+23& 0.78E+04& 0.54E+03& 0.14E+03&  3.73&  0.18&  0.14&  0.73\\
G24.6 %& 24.63& 0.18& 116& 7.3 &  4.15&  3.20& 10.00& 30.00
& 0.60E+22& 0.47E+03& 0.19E+04& 0.68E+03&  2.73&  0.74&  0.28&  0.37\\
G24.8 %& 24.80& 0.10& 111& 7.3 &  5.13&  7.00& 17.00& 160.00
& 0.32E+23& 0.20E+04& 0.15E+05& 0.40E+04&  3.77&  0.36&  0.18&  0.40\\ 
\hline 
\end{tabular} 
\label{tab2} 
\end{center}  
	\end{table*} 
  
%%%%%%%%%%%%%%%%%%%%%%%%%%%%%%%%%%%%%%%%%%%%%%%%%%%%

	\begin{table*}  

\caption{Tails of ET and GETs.}

  \begin{center}
\begin{tabular}{llllllllllllllllllll} 
\hline 
\hline 
Name %& $l$& $b$ &$\vlsr$& $d$ 
& Length $X$ &Width $Y$ & ${dv \/dx} \sin\ i$ 
& $\langle \Ico \rangle$ & $\langle N_{\rm H_2}\rangle$ 
& $\langle n_{\rm H_2}\rangle$ & $M_{\rm mol}$ \\ 

%& deg & deg & \kms & kpc
& pc &pc& \kms pc$^{-1}$ 
& K \kms & \htwocol 
& \htwocc & $\Msun$ \\ 

 \hline 
 
M16 E %& 17.04& 0.75& 25& 2 
& $\sim 4$ & $\sim 0.4$ & $-0.3$ 
&$\sim 30$ & $\sim 6$E21 
& $\sim 5$E3 & $\sim 230$ \\

M16 W %& 16.96& 0.78& 25& 2 
&$\sim 6$ & $\sim 0.4$ & $\sim \pm 1$ wavy
&$\sim 20$ & $\sim 4$E21
&$\sim 3$E3 & $\sim 210$ \\
\hline 
G31 %& 30.99& -0.05& 83& 5.5 
& $\sim 21$ & $\sim 2.5$& $+0.2$ 
&$\sim 60$ & $\sim 1.2$E22
& $\sim 1.6$E3 & $\sim 1.4$E3\\

G24.6% & 24.63& 0.18& 116& 7.3 
& $\sim 25$ & $\sim 3.6$& $+0.03$ 
&$\sim 15$ & $\sim 3$E21
& $\sim 270$ & $\sim 6$E3\\

G24.8 % & 24.80& 0.10& 111& 7.3 
& $\sim 50$ & $\sim 19$& $-0.09$ 
&$\sim 30$ & $\sim 6$E21
& $\sim 100$ & $\sim 1.2$E5\\

\hline 

\end{tabular} 
\label{tab1} 
\end{center}  
\end{table*}

  \section{Giant Elephant Trunks}
\subsection{GET G31-0.05+83} 

A large-sized ET-like molecular object, hereafter giant elephant trunk (GET), with the head clump position at G30.99-0.05+82.5 ($l=30\deg.99,\ b=-0\deg.05, \ \vlsr=83$ \kms, abbreviated as GET G31) was found during our \co-line study of the galactic molecular bow shock at G30.5 associated with the SF complex W43 (Sofue et al. 2018). W43 and associated giant molecular clouds (GMC) are located in the tangential direction of the 4-kpc molecular arm at a distance of 5.5 kpc. We here assume that the GET is at the same distance because of the close radial velocity.

\begin{figure*} 
\begin{center} 
\includegraphics[width=12cm]{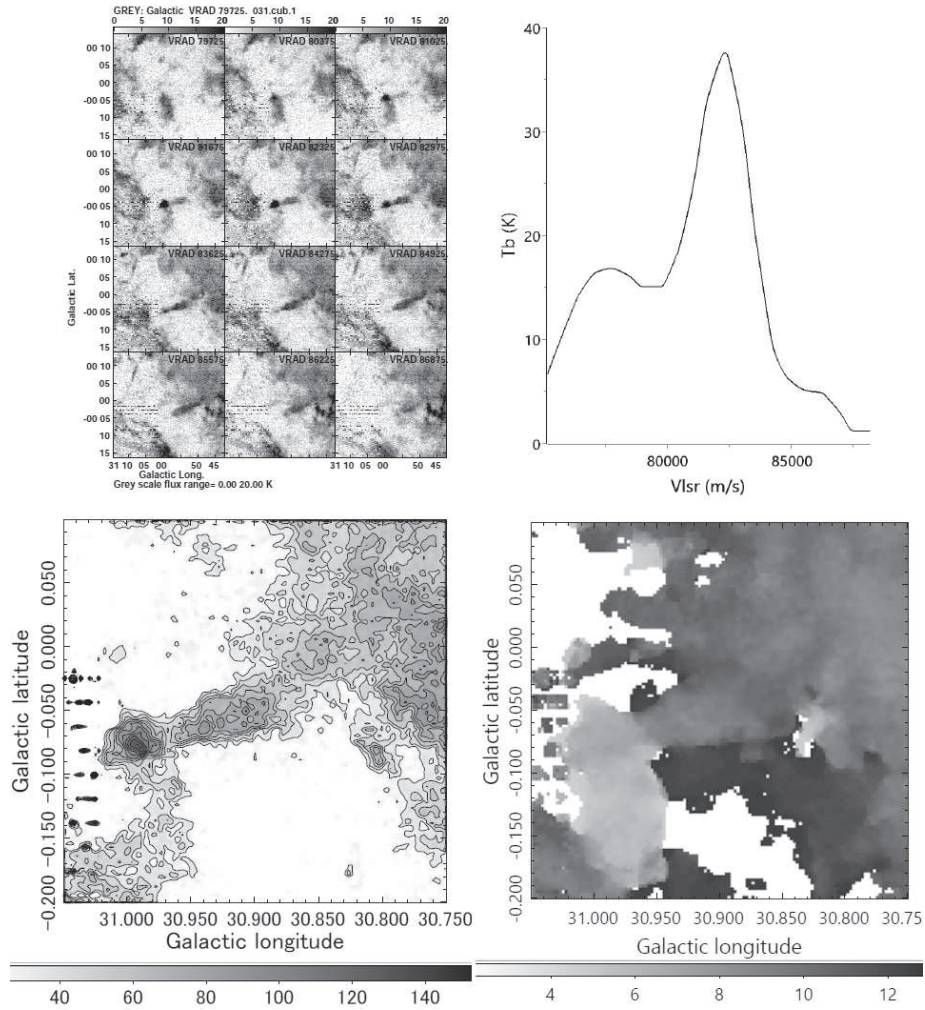} 
\end{center}
\caption{GET G31 in \co line: (tl) Channel maps of $\Tb$, (tr) line profile at the head clump, (bl) Integrated intensity $\Ico$ (contours start at 30 with interval of 10 \Kkms), and (br) velocity field (moment 1). (tl=top left, tr=top right, bl=bottom left, br=bottom rgith)} 
\label{G31} 
 	\end{figure*}

Figures \ref{G31} shows the obtained maps for GET G31. Although we examined infrared maps from ATLASGAL, we could not find any clear corresponding features.
GET G31 shows up in the channel maps of the \co-line brightness at around $\vlsr \sim 83$ \kms. A bright head clump at the eastern end of the structure is tailing toward the west, and the tail merges with the large molecular complex surrounding the SF site W43. The integrated intensity map shows that the GET is composed of a dense and compact head clump followed by a tail extending toward the west. The total length is about $X\sim 0\deg.2=20$ pc, and the full width of the tail is about $Y\sim 0\deg.02=2$ pc.

The head clump has peak brightness as high as $\Tb=40$ K, integrated intensity $\Ico=120$ K \kms, and velocity width $\delta v \simeq 3.5$ \kms. The full width of half maximum, or the size diameter, after beam correction is measured to be $D=1.0$ pc. 
These lead to \htwo column density of 
$N_{\rm H_2}=2.4\times 10^{22}$ \htwocol, and volume density 
$n_{\rm H_2}\sim 7.8 \times 10^3$ \htwocc.
The total mass of molecular gas is estimated to be 
$M_{\rm mol}\sim 540 \Msun$. This is about four times the the virial mass $M_{\rm vir}\sim 140 \Msun$. 
Thus the clump is gravitationally stable. 
 
The Jeans time in the clump calculated for the volume density of the molecular gas is on the order of 
$t_{\rm J}\sim \sqrt{1/4\pi G\rho}\sim 0.18$ My, where 
$\rho=\mu \mH n_{\rm H_2}$. 

Because the \co line is optically thick, the brightness temperature may approximately represent the excitation temperature, so that the sound velocity is related to $\Tb$ through $c_{\rm s} = \sqrt{\gamma RT}$ with $\gamma=5/3$.
This leads to Jeans length and mass of 
$\lambda_{\rm J}\sim 0.14$ pc and 
$M_{\rm J}=4 \pi/3 (\lambda_{\rm J}/2)^3 \mu \mH n_{\rm H_2} \sim 0.73\Msun$. 
Thus, the head clump can be a forming site of a gravitationally bound cluster of low-mass stars, possibly a birth place of a globular cluster.

The offset-velocity field (moment 1 map) as well as the channel maps show a velocity gradient at $dv/dx\ \sin\ i \sim +0.2$ \kms pc$^{-1}$ along the major axis, showing that the tail is receding from the head. This means that the head is on the near side, if the GET is accelerated from the head toward tail. 

\subsection{GET G24.6+0.2+115}

\begin{figure*} 
\begin{center} 
\includegraphics[width=12cm]{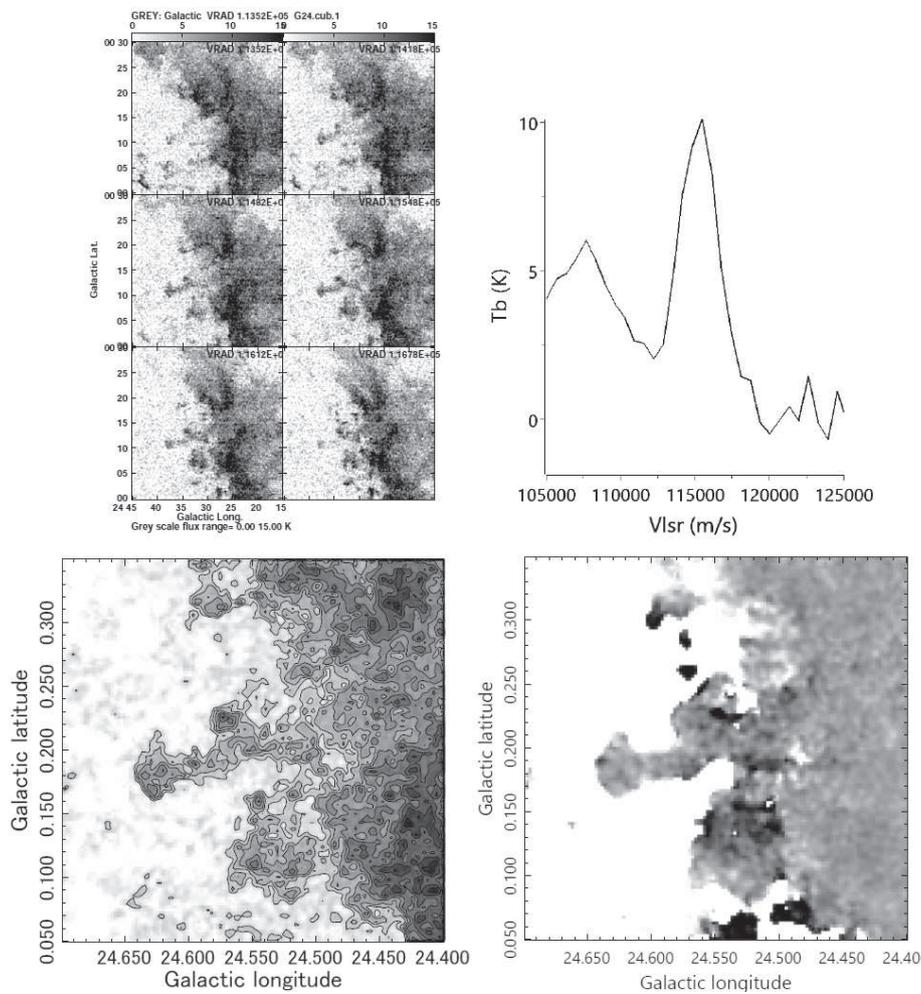} 
\end{center}
\caption{GET G24.5 in \co line: (tl) Channel maps of $\Tb$, (tr) line profile at the head clump, (bl) Integrated intensit $\Ico$ (contours start at 10 with interval 5 \Kkms), and (br) velocity field (moment 1). } 
\label{ET24W} 
 	\end{figure*}

During a study of the galactic shock properties toward the tangential direction of the 3-kpc (Norma) arm, we noticed wavy structures at the eastern sharp edge of a GMC facing the HII regions at G24 (Sofue 2019). Among the waves, we here focus on a GET shown in figure \ref{ET24W}. The head is located at G24.6+0.2+115 ($l=24\deg.63,\ b=+0\deg.18, \ \vlsr=115$ \kms; GET G24.6). The distance of the GMC was determined to be 7.3 kpc as the tangent point of the Norma Arm, and we adopt the same distance for the GET. 

GET G24.6 shows up as a horizontally extended ridge of \co brightness in the channel maps, emerging from the GMC on the right of figure \ref{ET24W}(a). 
The line profile indicates a peak brightness toward the head clump to be 10 K, and the line width of $\delta v=3.2$ \kms centered at $\vlsr=115$ \kms. 
The $\Ico$ map shows a head clump on the eastern end, followed by a cometary tail extending toward the west. The length of the tail is about $X\sim 0\deg.18=23$ pc and width $Y\sim 2.5$ pc. 

$\Ico$ at the head is measured to be $\Ico \simeq 30$ K \kms, leading to a column density of $N_{\rm H_2}=6\times 10^{21}$ \htwocol, volume density 
$n_{\rm H_2}\sim 4.7\times 10^2 $ \htwocc.
The molecular mass of the head clump is
$M_{\rm mol}=1.9 \times 10^3 \Msun$, which may be compared with the virial mass of 
$M_{\rm vir}\sim \sim 0.7 \times 10^3 \Msun$, indicating that the head is gravitationally stable.
 
The Jeans time, wave length, and mass are calculated to be 
$t_{\rm J}\sim 0.74$ My, 
$\lambda_{\rm J}\sim 0.28$ pc, and 
$M_{\rm J}=0.37 \Msun$. 

The offset-velocity field as well as the channel maps show a velocity gradient at $dv/dx\ \sin\ i \sim +0.04$ \kms pc$^{-1}$ along the major axis.
This positive gradient means that the head is on the near side of tail, if the GET is accelerated along the tail. 
 
\subsection{GET G24.8+0.1+111}

\begin{figure*} 
\begin{center} 
\includegraphics[width=12cm]{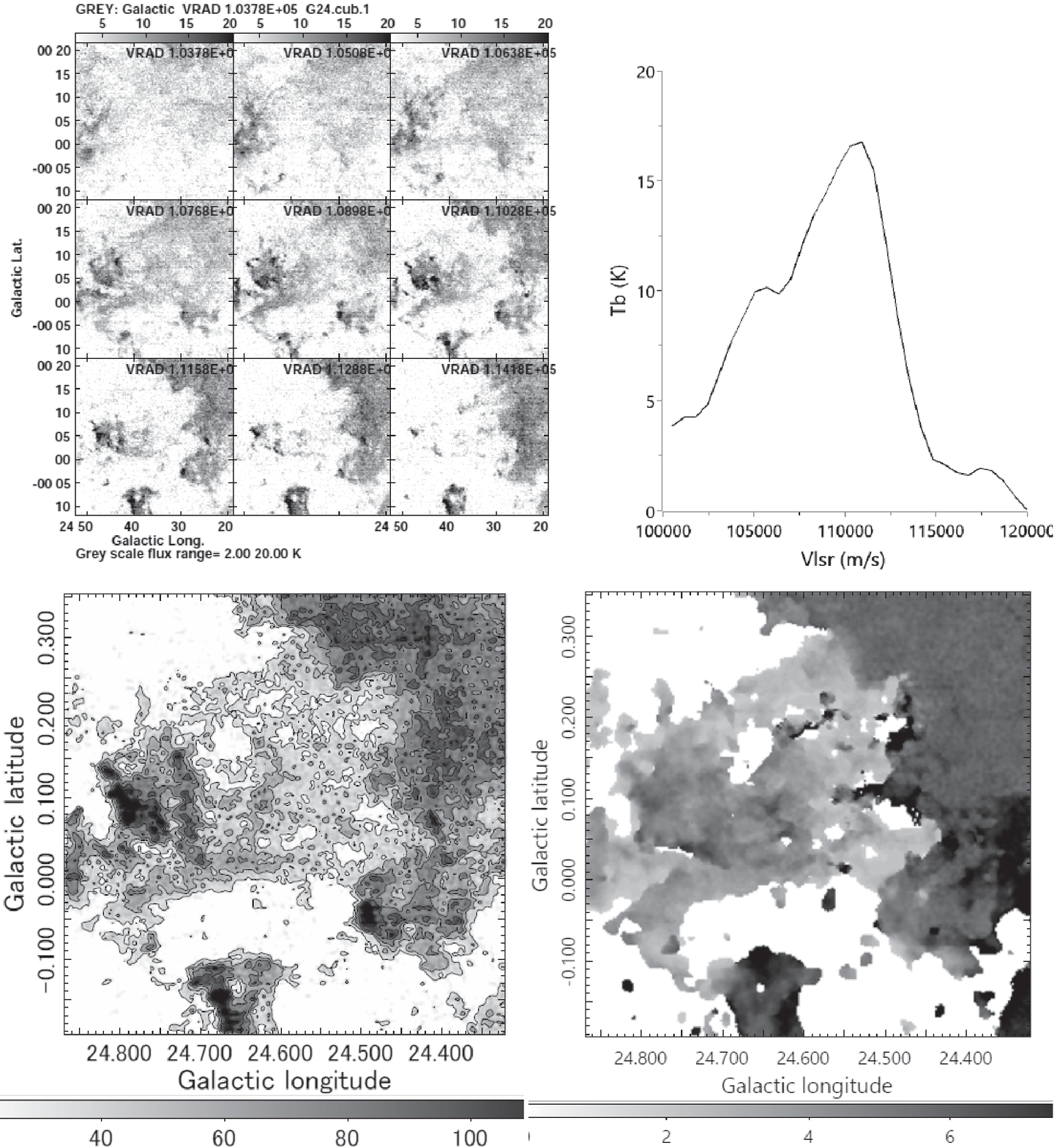} 
\end{center} 
\caption{GET G24.8 in \co line: (tl) Channel maps of $\Tb$, (tr) line profile at the head clump, (bl) Integrated intensit $\Ico$ (contours from 30 with interval 20 \Kkms), (br) velocity field (moment 1).} 
\label{ET24E} 
 	\end{figure*}

An order of magnitude larger scale GET is found at G24.8+0.1+111 ($l=24\deg.8,\ b=+0\deg.1, \ \vlsr=111$ \kms; GET G24.8) as shown in figure \ref{ET24E} the same distance of 7.3 kpc. This GET shows up as a broad tongue shape of \co brightness in the channel maps, emerging from the western GMC. The whole structure is clumpy both in space and velocity.
The tail has a dimension of $X\sim 50$ pc and $Y\sim 25$ pc, and is bifurcated to a few ridges.

The head clump is elongated perpendicular to the major axis by %10x30 pix
$D_x \times D_y \sim 3 \times 9$ pc, or $D\sim 5.1$ pc. 
The peak brightness toward the clump is $\Tb\simeq 17$ K, and the line width is $\delta v=7$ \kms. 

$\Ico$ at the head is as strong as $\Ico \simeq 160$ K \kms, and the column and volume densities are
$\NHtwo \sim 3\times 10^{22}$ \htwocol, and
$n_{\rm H_2}\sim 2\times 10^3 $ \htwocc.
Total molecular mass is estimated to be 
$M_{\rm mol}=1.5 \times 10^4 \Msun$, which is comparable to the virial mass of $M_{\rm vir}\sim 0.4\times 10^4 \Msun$, indicating that the clump is gravitationally stable.

The Jeans time, wave length, and mass are calculated to be 
$t_{\rm J}\sim 0.36$ My, 
$\lambda_{\rm J}\sim 0.18$ pc, and 
$M_{\rm J}=0.4 \Msun$. 

The offset-velocity field shows a velocity gradient at $dv/dx\ \sin\ i \sim -0.11$ \kms pc$^{-1}$ along the major axis.
The negative gradient suggests that the head is on the far side of tail.

\begin{figure} 
\begin{center}  
\includegraphics[width=8cm]{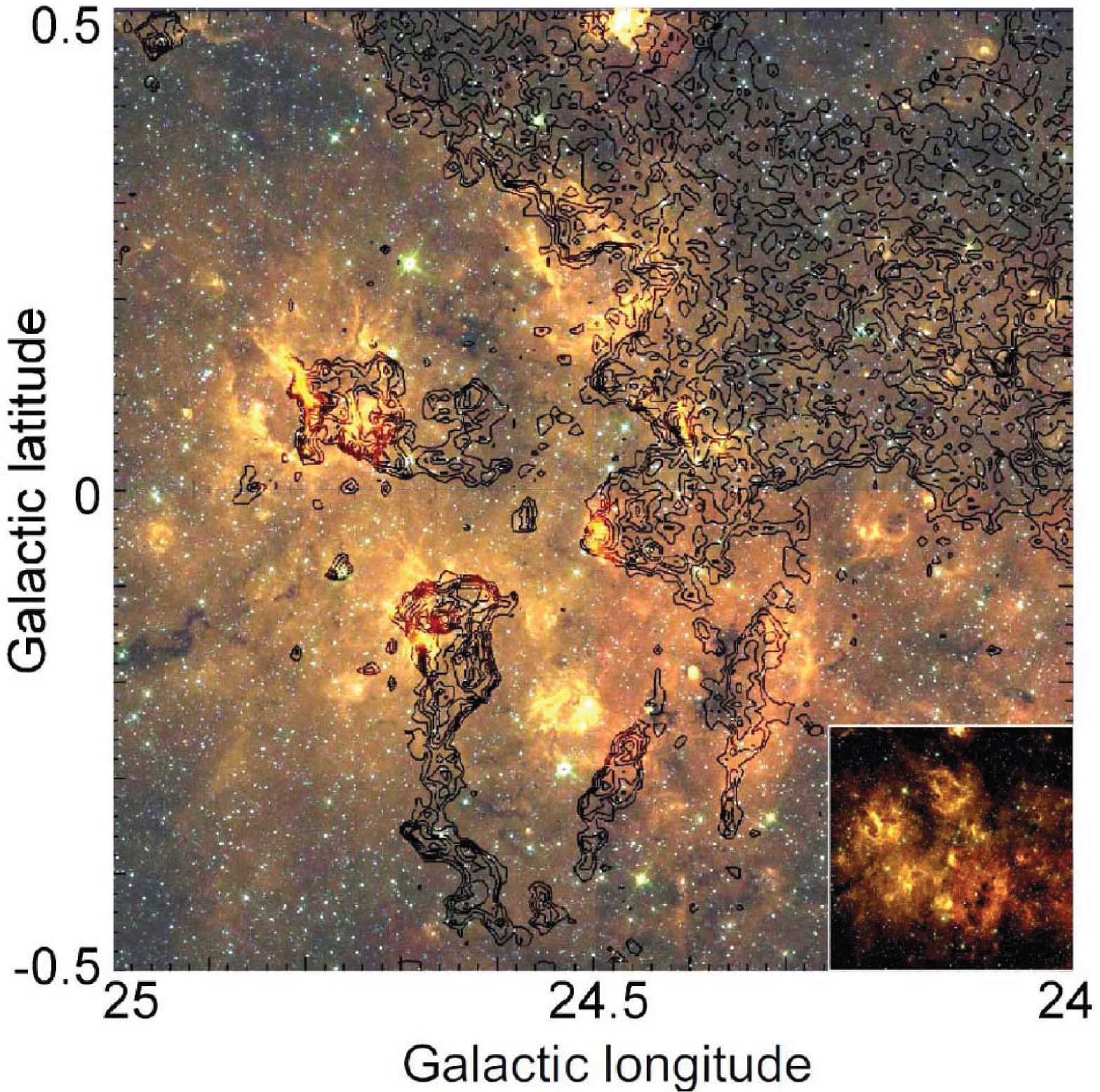} 
\end{center} 
\caption{
 Rayleigh-Taylor instability growing on the concave surface of a GMC at G24.5 on a CO channel map at 112 \kms (Sofue 2019). Overlaid is a composite color map of infrared emisstions at 4, 6, and 8\ \mum \ from ATLASGAL. The giant elephant trunk G24.8 is associated with a compact HII region at the eastern edge of the head clump.}   
\label{RTIG24} 
 	\end{figure}

This GET is associated with a compact HII region, as shown in a wider field map in figure \ref{RTIG24}, where a $\Tb$ channel map at 112 \kms by contours is overlaid on a composite colored infrared map (4, 6, 8 $\mu$m) from ATLASGAL. The molecular head clump is closely contacting the HII region at the eastern edge.

	\begin{figure} 
\begin{center} 
\includegraphics[width=7cm]{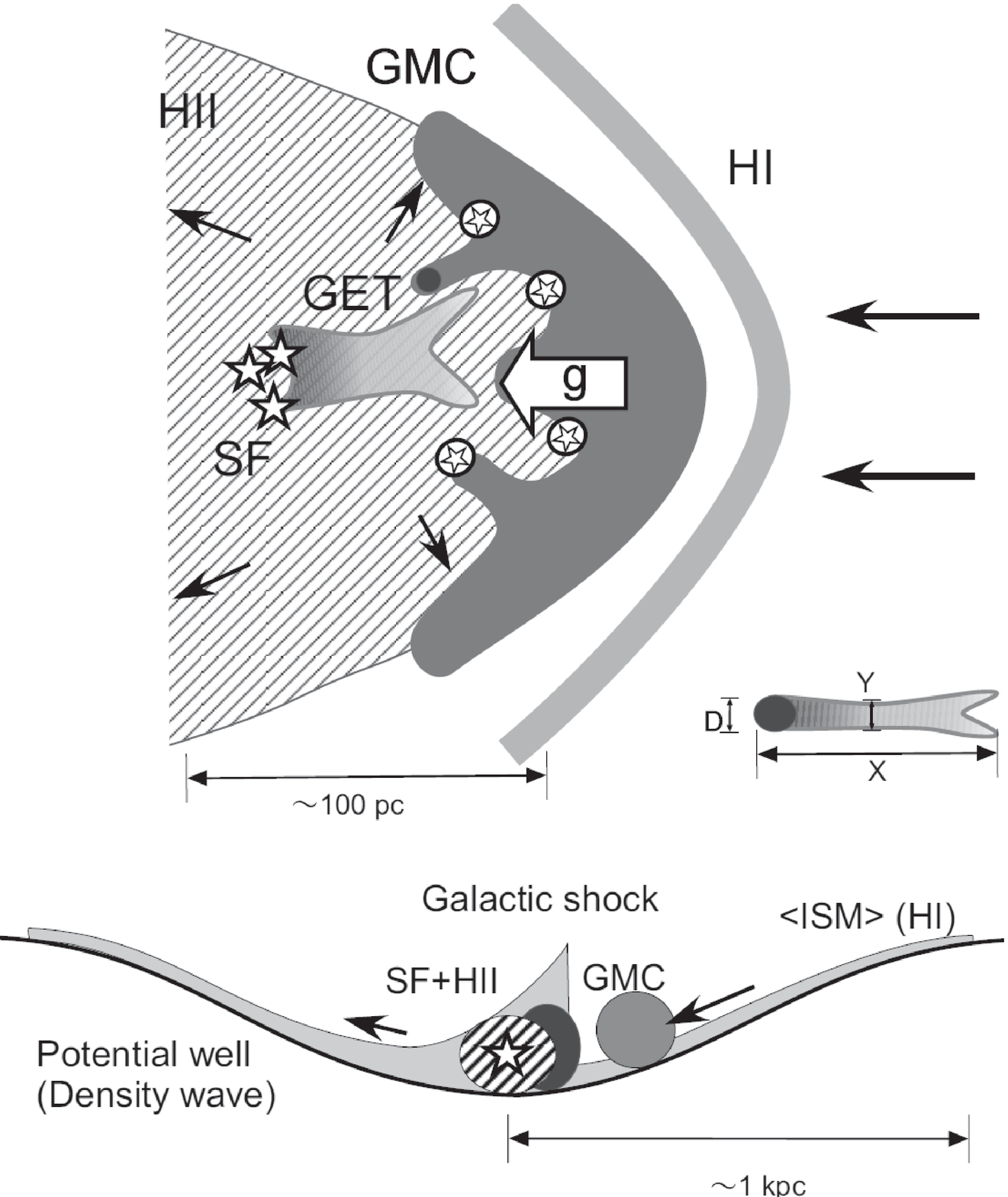} 
\end{center}
\caption{Schematic illustration of shock compression of GMC in the galactic shock in the potential well of a spiral density wave. Expanding HII gas from pre- existing SF site causes additional compression from the down-stream side, causing deceleration of GMC surface, which leads to the RT instability.} 
\label{illust} 
 	\end{figure}

  \section{Discussion and Summary}

We have found three giant elephant trunks (GET) emerging from the GMCs in the Scutum (G31) and Norma (G24) spiral arms. Derived paramters are listed in tables \ref{tab1} and \ref{tab2}. The GETs exhibit similar cometary morphology to the ETs in M16 and Rosette Nebulae. However, the sizes and masses are by an order of magnitude larger. The head clumps of the GETs are gravitationally stable in contrast to the ETs in M16, which are gravitationally marginal. 

Since the GETs are located in the direction of the tangent points of galactic rotation, the error in the kinematical distance is large compared to the region sizes. However, we consider that the GETs are physically associsted with the GMCs and HII regions for the following reasons, and discuss their properties and possible formation mechanism in the next subsections.
i) GETs are in touch with the GMCs on the sky,
ii) radial velocity is within GMC's velocity dispersion,
iii) trunks are protruding from the concave surface of the GMCs,
iv) the heads point the HII regions, as well as vi) the arm's gravity center,
and v) the tails are coherently parallel to the galactic plane. 

\subsection{Gravitationally bound GET heads}

GETs G24.6 and G31 have head clumps of diameter of a few pc and gas mass several hundred $\Msun$. The tails have width of $Y\sim 2-4 $ pc and length $X\sim 20-25$ pc. These two GET heads are not associated with 8 \mum emission, 
 indicating weaker UV illumination compared to M16 and G24.8.  

GET G24.8 is an order of magnitude more massive than the other two GETs, and is associated with a compact HII region at the eastern edge of the head clump. The head clump is as massive as $\sim 1.5 \times 10^4 \Msun$, significantly greater than the virial mass, and hence the clump is gravitationally bound. The tail is broad and extended for $Y\sim 20$ pc, and the length is as large as $X\sim 50$ pc. The total mass including the tail is as large as $\sim 10^5 \Msun$.

The Jeans masses in the head clumps of GETs have been approximately estimated by assuming that the kinetic temperature is equal to the observed brightness temperature, because the \co line is optically thick. The masses are found to be sub-solar.
The head clumps might be, therefore, forming sites of gravitationally stable clusters of low-mass stars, possiblly a new type of globular culsters of population I, which would be of low luminosity.

\subsection{Coherent emergence of GET in the spiral arm}

A remarkable feature in the three GETs is their coherently horizontal head-tail structures. They all extend from the parent GMC on the west toward the head clump in the east, parallel to the Galactic plane. Therefore, the GETs are stretched from the up-stream side toward down stream of the gas flow in the spiral arms. 

It is also emphasized that the GMCs are coherently concave to the east, which has been argued to be a result of bow-shock in the galactic shock wave by a supersonic flow from the west to east (Sofue et al. 2018; Sofue 2019a, b). 

These particular alignment of the GET and GMC suggests their common origin related to the galactic-shocked GMC in the density wave of spiral arms, as illustrated in figure \ref{illust}. Below, we model this idea based on a Rayleigh-Taylor instability.
 
\subsection{Rayleigh-Taylor instability on GMC surface}

\def\nHII{n_{\rm HII}} \def\nHtwo{n_{\rm H_2}} 

In our recent paper (Sofue 2019), we showed that the shock-compressed inner molecular edge of a GMC in G24 region encountering the spiral density wave of the Norma arm suffers from Rayleigh-Taylor instability (RTI) due to the deceleration of high-density molecular gas by low-density HII gas.  
Figure \ref{RTIG24} shows the G24.5 region in CO at 112 \kms by contours overlaid on infrared composite image (4\mum, 6\mum, 8\mum) taken from the ATLASGAL. The GMC on the right side is concave to the HII regions, and the eastern surface is sharp and wavy with periodical appearence of bay-and-peninsula structure with each peninsula appearing to develop to a GET.

The RTI has been modeled for ETs in local HII regions (Frieman 1954; Spitzer 1954; Osterbrock 1957), and recently by numerical simulations of the ionization front at the molecular clouds ({Whalen \& Norman} {2008}; {Mackey \& Lim} {2010}; {Mackey \& Lim} {2010}; {Chauhan et al. 2011}). However, the GETs found here may need some larger-scale cause of the acceleration on the GMC. We consider a possible mechanism of RTI in the spiral arms.

The wavelength $\lambda$ and time scale $\tau$ of RTI is related to the acceleration $g$ of gas at the front through (Frieman 1954)
\be
\lambda \sim 2 \pi g \tau^2 .
\label{lambda}
\ee
Approximating the tongue's extent $x$ by the RTI wave length $\lambda$, the velocity $v$ of the tail with respect to the GET head is related to $g$ as
\be
v\sim \sqrt{2g x},
\ee
or 
\be
dv/dx \sim \sqrt{g/2x}.
\label{dvdx}
\ee 
The growth time of the instability is given by
\be
\tau \sim \sqrt{x/2\pi g}.
\ee

\subsection{Cloud surface deceleration and RTI}

As to the cause of deceleration $g$, we may consider two mechanisms: (i) acceleration by the density wave potential $g_1$, and (ii) that by encounter with expanding HII front $g_2$. In the present case, both act in the same direction, so that the net acceleration is given by $g\sim g_1+g_2$. 
 
(i) Consider a GMC falling down along the potential slope of the density wave and encounter the pre- existing low-density interstellar gas stagnated in the valley of the potential well (figure \ref{illust}). 
 
The deceleration $g_1$ is, then, approximated by
\be
g_1 \sim { \delta \phi \/\Lambda } \sim {\epsilon \Vrot^2 \ \sin\ p\/ 2 \Lambda},
\ee 
where $\delta \phi \sim \epsilon \Vrot^2/2$ is the density wave potential and $\Lambda$ is the width of gaseous shocked lane, within which the cloud is stopped, and $p$ is the pitch angle of the spiral flow. 
As typical parameters for the density wave, we may take $\epsilon \sim 10$\%, $\Vrot\sim 200$ \kms, and $\Lambda \sim 100$ pc, and $p\sim 12\deg$. Then we have $g_1 \sim 1.3\times 10^{-8}$ cm s$^{-2}$.

(ii) RTI is a classical mechanism considered for ET formation at the interface of an HII region and surrounding dense gas. Deceleration of a molecular cloud surface pushed by HII gas is given by 
\be
g_2\sim {1\/ \mu \mH n_{\rm H_2}} {dp\/dx},
\ee
where $dp/dx$ is the pressure gradient, and can be approximated by 
$dp/dx \sim (p_{\rm HII}-p_{\rm GMC})/\Delta \sim p_{\rm HII}/\Delta$. 
The former term represents the dynamical pressure due to the internal pressure of HII gas, and the latter molecular gas pressure, where the former pressure is much higher than the latter. Rewriting the pressure of the HII gas by sound velocity $c_{\rm s}$ and density, we obtain
\be
g_2\sim {c_{\rm s}^2\/ \Delta} {\nHII \/\mu \nHtwo}.
\ee
Taking $c_{\rm s}\sim 10$ \kms, $\Delta\sim 1$ pc, $\nHII \sim 10^2$ H cm$^{-3}$ and $\nHtwo \sim 3\times 10^3$ \Htwo cm$^{-3}$, we obtain
$g_2\sim 0.1 \times 10^{-8}$ cm s$^{-2}$, an order of magnitude smaller than $g_1$. 

\subsection{Arm-scale RTI}

We may, thus, consider that the density wave acceleration is the dominant source for the growth of GETs in G31 and G24 regions. Moreover, if there exists a bar in the inner Milky Way, the potential will be deeper and $\sin\ p$ may be larger than those assumed above for a normal spiral arm. So, we will here take $g\sim 2\times 10^{-8}$ cm s$^{-1}$, although its precise determination is a subject for the future.

Let us assume that the GET's length is comparable to the RT wavelength, $\lambda \sim X$, then the growth time of RTI of length $\sim 20$ pc is estimated to be 
$\tau \sim 0.7$ My. 
The velocity gradient is estimated to be 
$dv/dx\sim 0.4$ \kms pc$^{-1}$.

Considering the projection effect of the GET axes, this velocity gradient is consistent with the observed gradients of $(dv/dx)\sin\ i \sim -0.1$ to $0.2$ \kms pc$^{-1}$ in table \ref{tab1}. Positive gradients for G31 and G24.6 indicate positive $i$, so that the the heads are nearer and tails are stretched away from the Sun at $i\sim 30\deg$ and $\sim 4\deg$, respectively. On the other hand, GET G24.8 has negative gradient, indicating that the head is in the far side, and the tail is extending toward the Sun at $i\sim -14\deg$. 

\subsection{Galactic-scale GET}

We finally comment on the implication of the presently reported GETs not only for the interstellar physics but also for galactic dynamics of spiral arms. 

A large number of bow shocks of molecular gas concave to OB clusters and HII regions have been found in the SF-active arms of the barred spiral galaxy M83 (Sofue 2018). The bow-shaped molecular-HII associations exhibit remarkable similarity to those found in the Milky Way at G24 (Sofue 2019) and G31 (Sofue et al. 2018). Thereby, we suggested RTI as a formation mechanism of wavy structure of the bow surface on the concave side. 

Much greater GETs of sizes from $\sim 200$ pc to 1 kpc, or mammoth trunks, have been found in the central dark ring of an elliptical galaxy (Carlqvist, Kristen, and Gahm 1998), while no signature of SF activity is reported. Magnetized filaments and/or RTI by the galactic wind have been suggested for their origin.
 
Thus far, the GETs are not particular objects in G30 and G24 regions, but they may be a universal trunk phenomenon from $\sim 20$ pc to $\sim 1$ kpc scales, growing in spiral arms and galactic rings. Given the RTI origin is the case, such quantities like the size, mass, and velocity gradient would give useful information to look insights into the physics of galactic shock waves and rings.

\subsection{Summary}

We discovered three giant elephant trunks (GET) of molecular gas in the \co line archival data from FUGIN survey with the Nobeyama 45-m telescope at resolutions of 0.65 \kms and $20''$, or 0.5 pc at G30 and 0.7 pc at G24. Observed quantities and derived physical parameters are listed in tables \ref{tab1} and \ref{tab2}. The sizes and masses of the GETs are an order of magnitude greater than those of the ETs in the local HII regions such as those in M16.

GETs have a cometary structure, which is coherently aligned parallel to the galactic plane, pointing down-stream direction of the galactic flow, emerging from the surfaces of bow-shocked concave GMCs located in the west. 
The head clumps of GETs have masses $\sim 10^3 -10^4 \Msun$, about three times the virial masses, showing that they are tightly bound gravitationally. The Jeans masses calculated for the kinetic temperature assumed to be equal to the brightness temperature of the optically thick \co line are commonly sub-solar. We therefore suggest that the GET heads are possible sites for formation of globular clusters of low-mass stars.
We also point out that the GETs may be useful to probe not only the SF activity, but also the potential depth of the spiral arm and internal structure of galactic shock waves.

  \vskip 2mm
%%%%%%%%%%%%%%%%%%%%%%%%%%%%%% 
{\bf Aknowledgements}
The author is indebted to the authors of the survey data, particularly Prof. T. Umemoto of NAOJ and collaborators for the FUGIN CO survey data obtained with the Nobeyama 45-m telescope. The 8 \mum data were taken from the ATLASGAL data archives based on the Spitzer Telescope infrared survey. Data analysis was carried out at the Astronomy Data Center of the National Astronomical Observatory of Japan. 
%%%%%%%%%%%%%%%%%%%%%%%%%%%%

\begin{appendix} 

\section{Pressing and background-filtering methods}
\label{BGF}

Here is a brief description of the advanced pressing and background filtering (BGF) methods, originally developed for radio continuum scanning observations with a single-dish telescope (Sofue and Reich 1979). The pressing method is used to correct for the scanning effects, or stripes on the map, arising from variations the gain, atmospheric attenuation and emission, ground emission from side lobes, zero-level fluctuation, etc.. 

The BGF method removes background emissions such as the Galactic disc with large-scale intensity gradients, and abstracts embedded discrete as well as extended radio sources. This is similar to unsharp masking, but the result can better be used for quantitative analyses. The methods employ the following procedures.

\subsection{Pressing method}  
\begin{figure} 
\begin{center}   
(a)\includegraphics[width=6cm]{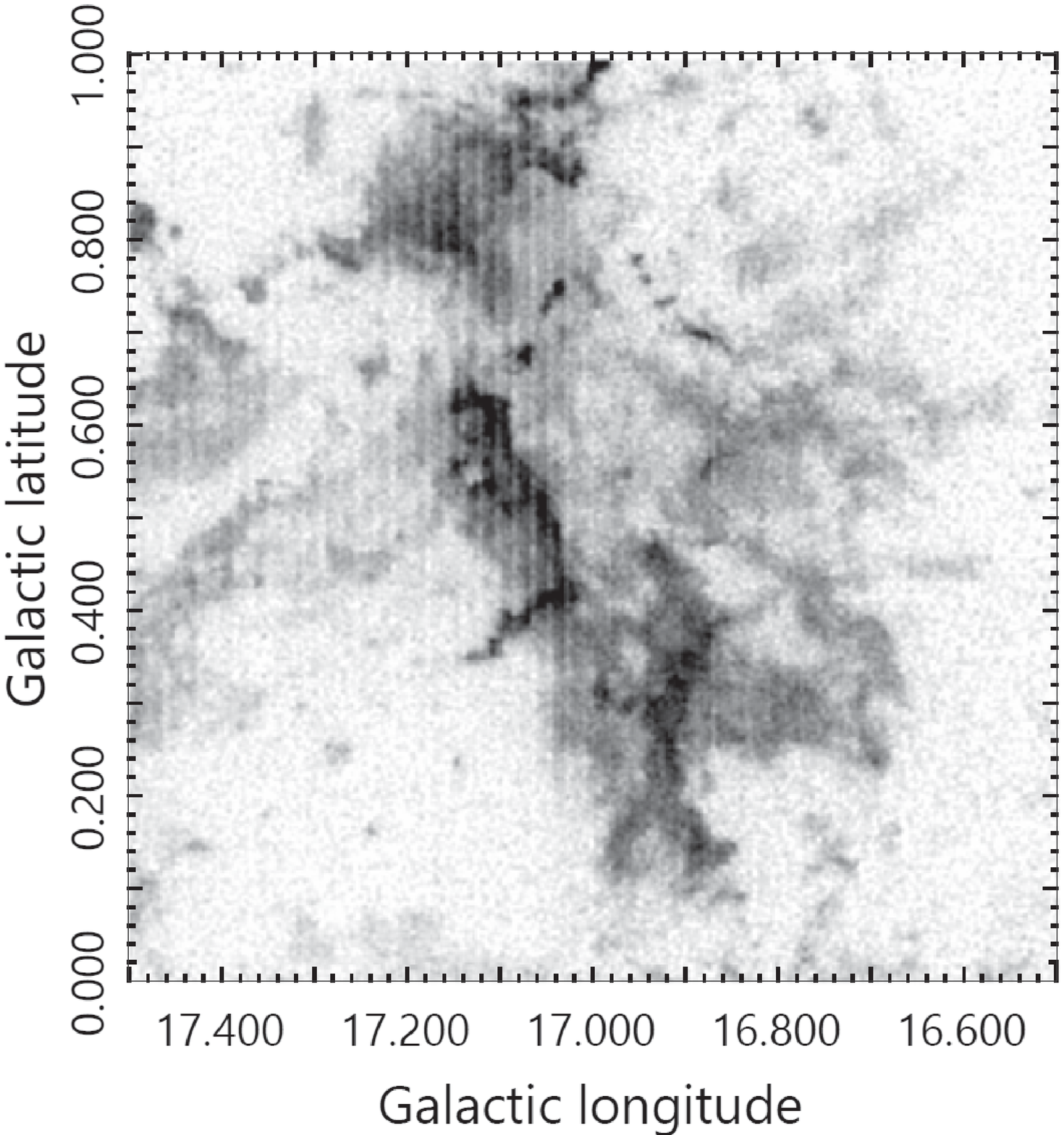}  
(b)\includegraphics[width=6cm]{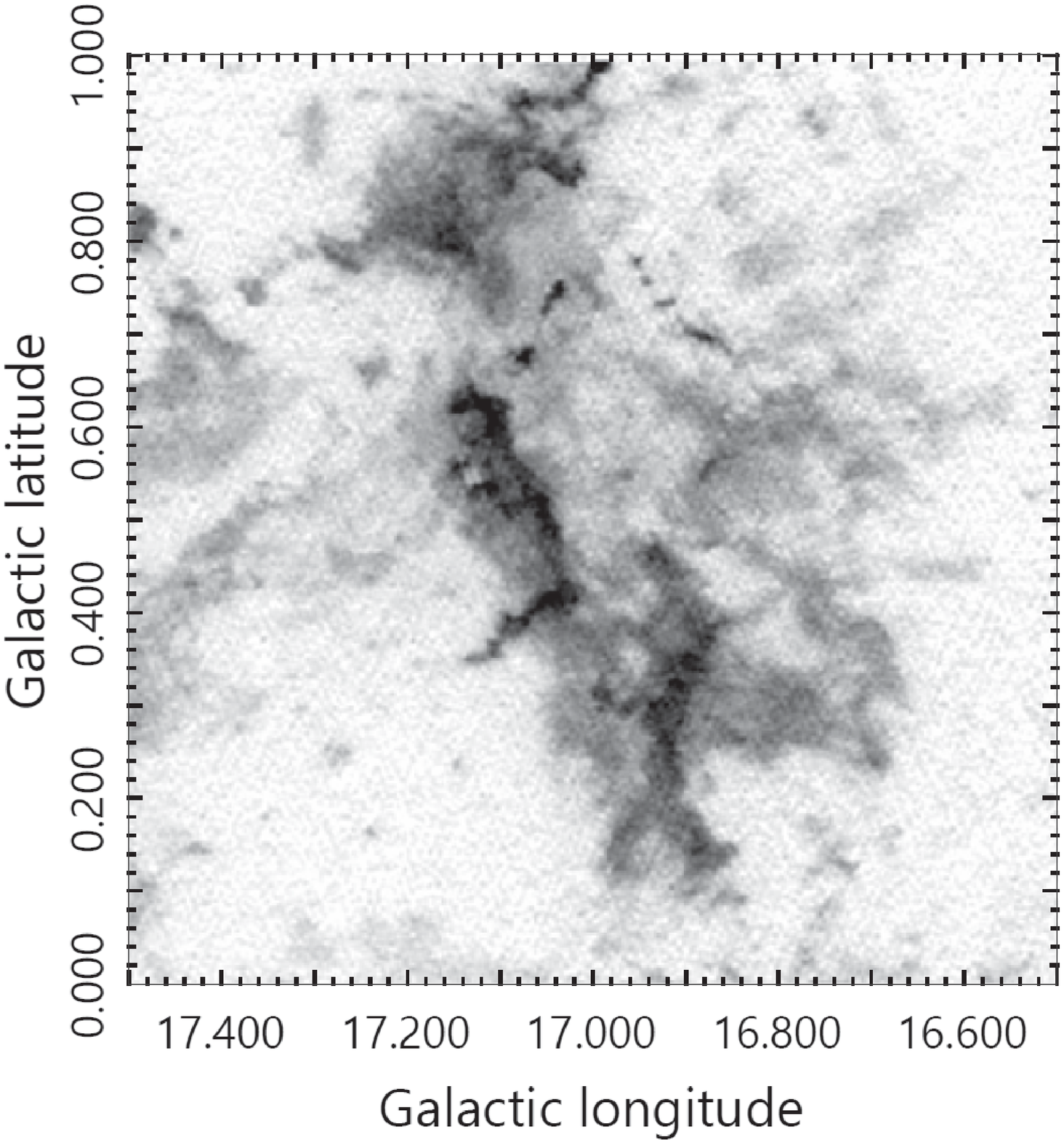}  
\end{center}
\caption{Example of application of pressing to Y-scan CO intensity map around M16 (a) before and (b) after correction.}   
\label{press} 
 	\end{figure}

Suppose that the original map was obtained by scanning in the X direction, and Y is the direction perpendicular to X.
\begin{itemize}
\item Map A is smoothed in Y direction to get a perp-smoothed map B by a Gaussian or a box beam with (X, Y) widths of (1,$p$) pixels, where $p\sim 10-20$ pix., depending on the nature of the effect. 
\item B is subtracted from A to get scan effect C=A-B. 
\item C is smoothed in X direction by a Gaussian or box beam of width $(q,1)$ pix. to get smooth scan effect D, where $q=20-50$ in the present case with the map dimension of $848 \times 848$ pix., but depends on the scale length of variation along the scan. Instead, one may fit each scan effect by a polynomial or synusoidal function of X as in Sofue and Reich (1979).
\item D is smoothed in Y direction to get sub-smoothed map E by  a bem of width (1,$r$) with $r=10-20$ in order to recover artificial flux increase or decrease.
\item E is subtracted from A to get the pressed map F=A-E. 
\end{itemize}
Depnding on the remaining effect, apply the same to the result again or more times. If scan direction is in the Y direction, apply the same by reversing X and Y. 
Figure \ref{press} shows an example of the application to the FUGIN \co 
map around M16 at $\vlsr=24.5$ \kms. 

\subsection{BGF method} 
\begin{itemize}
\item The original map A is smoothed to yield a smoothed map B by a Gaussian beam of representitatve width for the desired object sizes to be abstracted.
\item B is subtracted from A to get source C=A-B. 
\item Negative C pixels are replaced with 0 to get source above zero D.
\item D is subtracted from A to get backgound E=A-D.
\item E is smoothed to get smooth background F.
\item F is subtracted from A to get source G=A-F.
\item Negative G pixels are replaced with 0 to get source H.
\item H is subtracted from A to get smooth background I.
\item I is smoothed to get smoothed background J.
\item These are repeated until J gets stable (2-3 times). 
\item Finally, K=A-J gives the BGF map, and J is the background. 
\end{itemize}
 
\end{appendix}

\end{document}